\documentclass[conference]{IEEEtran}
\IEEEoverridecommandlockouts
\usepackage{cite}
\usepackage[tbtags]{amsmath}
\usepackage{amssymb,amsfonts}
\usepackage{algorithmic}
\usepackage{graphicx}
\usepackage{textcomp}
\usepackage{xcolor}
\usepackage{multirow} 
\usepackage{comment}

\newcommand\numberthis{\addtocounter{equation}{1}\tag{\theequation}}

\def\BibTeX{{\rm B\kern-.05em{\sc i\kern-.025em b}\kern-.08em
    T\kern-.1667em\lower.7ex\hbox{E}\kern-.125emX}}
\begin{document}

\title{Modelling Micro-Doppler Signature of Drone Propellers in Distributed ISAC}

\author{\IEEEauthorblockN{
Heraldo Cesar Alves Costa\IEEEauthorrefmark{1},   
Saw James Myint\IEEEauthorrefmark{1},   
Carsten Andrich\IEEEauthorrefmark{1},
Sebastian W. Giehl\IEEEauthorrefmark{1}, \\
Christian Schneider\IEEEauthorrefmark{1}\IEEEauthorrefmark{2},  
Reiner S. Thom\"a\IEEEauthorrefmark{1}   
}   

\IEEEauthorblockA{\IEEEauthorrefmark{1}
Institute for Information Technology and Thuringian Center of Innovation in Mobility,\\
Technische Universität Ilmenau, Ilmenau, Germany}
\IEEEauthorblockA{\IEEEauthorrefmark{2}
Fraunhofer Institute for Integrated Circuits IIS, Ilmenau, Germany}
{heraldo-cesar.alves-costa@tu-ilmenau.de}
}

\IEEEoverridecommandlockouts
\IEEEpubid{
\makebox[\textwidth]{\raisebox{-2ex} {\scriptsize 979-8-3503-2920-9/24/\$31.00~\copyright2024 IEEE | DOI: 10.1109/RADARCONF2458775.2024.10548468} \hfill}}

\maketitle


\begin{abstract}
    Integrated Sensing and Communication (ISAC) comprises detection and analysis of non-cooperative targets by exploiting the resources of the mobile radio system. In this context, micro-Doppler is of great importance for target classification, in order to distinguish objects with local movements. For developing algorithms for target classification, it is necessary to have a large amount of target signatures. Aiming to generate these data, this paper proposes a mathematical model for the micro-Doppler of drone rotating propellers, and validate the proposed model by comparing it to measured micro-Doppler. Results show that the proposed mathematical model can generate micro-Doppler data very similar to those from measurement data. 
\end{abstract}

\begin{IEEEkeywords}
ISAC, BiRa, dynamic reflectivity, drone, micro-Doppler, bistatic, OFDM.
\end{IEEEkeywords}

\section{Introduction}\label{section:Introduction}
Integrated Sensing and Communication (ISAC) is considered to be one of the key features of future 6G mobile radio. 
ISAC stands for the detection and analysis of non-cooperative objects (“targets”) by exploiting the inherent resources of the mobile radio system on both the radio access and network level, using the signals transmitted simultaneously for both communication and target illumination purposes.
Therefore, ISAC performs target detection in configurations typical for communication systems, which comprise the waveform — usually Orthogonal Frequency Division Multiplexing (OFDM) and derivatives —, its numerology, multiuser access (OFDMA, TDMA), pilot schemes, multi-link (multistatic) propagation, channel state estimation, and synchronization \cite{Thoma2}.

One important aspect of multi-link propagation characterization is bistatic target dynamic reflectivity and, in this context, micro-Doppler, which refers to the small fluctuations in the Doppler shift caused by the motion of a target's internal parts, such as rotating blades or moving limbs. Bistatic case is the native geometry in mobile radio, and micro-Doppler provides valuable information about the target's moving components, which can be used to identify the type of target and distinguish it from other objects. Due to these reasons, micro-Doppler is a phenomenon that is becoming increasingly important in the context of ISAC systems.

Bistatic micro-Doppler measurement and analysis are important tools for characterizing targets, since bistatic geometry provides diversity advantage over monostatic case, giving additional information about the target's motion and structure \cite{Smith2008radar}.
Correct interpretation of bistatic micro-Doppler signatures can be critical for accurate target tracking and classification, especially in complex environments where multiple targets may be present.
Therefore, the development of advanced signal processing techniques for bistatic micro-Doppler data is of great interest to the radar community.
Nevertheless, despite the importance of bistatic micro-Doppler measurements in ISAC systems, there is currently a limited amount of data and research on this topic compared to monostatic micro-Doppler. 

In order to fill this gap and develop adequate techniques, it is important to have access to a large amount of target data.
\underline{Bi}statische-\underline{Ra}dar-Messeinrichtung (BiRa) measurement facility, installed at the Ilmenau University of Technology, can play an interesting role in the way to solve this problem, since it is able to automatically perform multiple multiaspect micro-Doppler measurements along a defined set of bistatic geometry positions, with different parameters, and without user intervention \cite{Thoma2}. This allows collecting numerous measurement data in a single unmanned measurement run.

However, training some classification algorithms such as machine learning and deep learning layouts requires a huge amount of training data.
In this scenario, a simple model is of great help for generating large amounts of micro-Doppler data, since a model can be more general than measurements. It is able to generate data for all distances, which is not always feasible with measurements. Moreover, a model can be scalable, in effort and accuracy, in order to avoid overcomplexity. 

The present work proposes a new OFDM bistatic micro-Doppler mathematical model for drone rotating propellers, developed based on the classic model for monostatic radars presented in \cite{Jianjiang2005}, intended to become an interesting tool for future development of ISAC target classification algorithms.

Section II introduces the mathematical formulation of the proposed model.
Section III presents the BiRa measurement system and the measurement setup used in the current paper.
Section IV validates micro-Doppler signatures generated with the proposed model by comparing them to those obtained from the measurements.
The conclusion and outlook can be found in Section V.


\vspace{5pt}
\section{Propeller Micro-Doppler Model}\label{section:Model}
\subsection{Classic Propeller Micro-Doppler Model}\label{subsection:classic_model}

    Suppose that a radar transmits a signal described as:
            
    \begin{equation}\label{eq:base_radar_signal}
        x(t) = \gamma_0 \exp(j\omega_0t),
    \end{equation}
    where $\gamma_0$ is the amplitude of the transmitted signal, $\omega_0 = 2\pi f_0 = 2\pi  \frac{c}{\lambda_0}$, $f_0$ is the transmitter central frequency, $c$ is the light speed, and $\lambda_0$ is the wavelength corresponding to $f_0$.
        
    If a scatter point target P has a range $R_P^{(m)}(t)$ with respect to a monostatic radar, its returns to this radar can be expressed as \cite{Richards2014} 
    
    \begin{equation}\label{eq:sP}
        y_P(t) = \gamma_P(t) \exp \bigg\{ j\omega_0 \left[t-\frac{2 R_P^{(m)}(t)}{c}\right] \bigg\} + z(t),
    \end{equation}
    where $z(t)$ is the noise, and $\gamma_P(t)$ is the amplitude of the returned signal, given by \cite{braun2014ofdm}
    
    \begin{equation}\label{eq:radar_equation}
        \gamma_P(t) = \gamma_0 \sqrt{\frac{c \sigma_P}{(4\pi^3)R_P(t)^4 f_0^2}},
    \end{equation}
    $\sigma_P$ is the radar cross section (RCS) of the scatter point $P$.

    Since the wavelengths in the proposed ISAC systems (Table \ref{tab_MicroDoppler_measurement_setup}) are smaller than the dimensions of most drone structures, except for some micro and nano drones which are out of the scope of this study, the electromagnetic scattering of these targets behave similarly to what happens in the optical area. 
    
    As the interaction among different scattering centers is very weak, the scattering can be assumed to be a local linear process, i.e., the target returns in a given range resolution cell is the superposition (coherent sum) of each independent scattering center in that resolution cell at that instant \cite{Jianjiang2005}.

    In this case, the target’s total scattering echo is given by the coherent sum of all independent scattering centers at that instant. 
    Therefore, considering a blade $i$ as a line of infinite point scatters going from the center \textbf{O} until a point \textbf{P\textsubscript{tip}}, distant $l = L_B$ from the center \textbf{O}, \cite{misiurewicz1997analysis} derives a mathematical model for the monostatic slow-time returns of a rotor with $N_B$ rotating blades of length $L_B$ each:            
    \begin{equation}\label{eq:SN}
    \begin{split}
        \tilde{h}_i(t) &= \sum_{i=1}^{N_B} \bigg\{ \tilde{\gamma}_i(t) e^{\left[j\frac{\omega_0}{c}\left(-2R_O(t)+L_B \cos(\omega t+\varphi_B (i)) \cos(\beta)\right)\right]} \\& \cdot L \operatorname{sinc}\left[\frac{\omega_0 L_B}{c} \cos(\omega t+\varphi_B(i)) \cos(\beta)\right] \bigg\} + \tilde{z}(t),
    \end{split}
    \end{equation}    
    where $\beta$ is the elevation angle from the radar to the rotation center with respect to the rotation plane, and
     \begin{equation} \label{eq:phi_B_i}
        \varphi_B(i) = \varphi_0 + 2\pi\frac{i-1}{N_B} .
    \end{equation}  

    It is important to note that changes in $R_P^{(m)}(t)$ generate changes in the phase of the signal. If the target is moving, $R_P^{(m)}(t)$ corresponding to each symbol echo is changing. As a result of this, the phase term becomes a changing function on $R_P^{(m)}(t)$, and the Doppler frequency is given by the derivate of this phase with respect to the slow-time \cite{Qu2019}.

    Therefore, this model represents the monostatic slow-time response of a rotor with $N_b$ blades, when our radar/sensor uses low-resolution CW waveform. This model can be used with good level of success for cases with both low-resolution and narrowband.

\subsection{Bistatic Geometry}\label{subsection:geometry}

    In order to be able to derive a similar model for the broadband bistatic case, it is first necessary to have a geometrical model of the bistatic range of a single rotating point.
    
    \begin{figure}[t!]
        \centering
        \includegraphics[width=8cm]{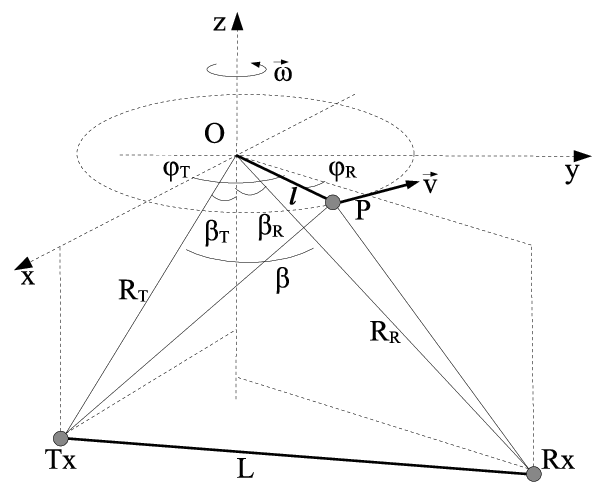}
        \caption{Bistatic rotating point geometry}
        \label{fig:rot_pt1}
    \end{figure}
    
    Such a math model was presented in \cite{Ai2011}, where the author develops the mathematical formulation of the bistatic range of a scatter point \textbf{P}, in a bistatic radar environment, rotating around a point \textbf{O}, according to Fig. \ref{fig:rot_pt1}. This bistatic range is
    \begin{equation} \label{eq:R_t}
        R_P^{(b)}(t) \approx R_O - A_B l \cos(\omega t+\varphi_B),
    \end{equation}
    where $R_O = R_T + R_R$ is the total bistatic range of the rotation center \textbf{O}, $A_B = \sqrt{4\cos(\beta/2)^2-(\cos\beta_T + \cos\beta_R)^2}$, and
     \begin{equation} \label{eq:phi_B}
        \varphi_B = \varphi_T - \arctan\left(\frac{\sin\beta_R \sin(\varphi_T - \varphi_R)}{\sin\beta_T + \sin\beta_R \cos(\varphi_T - \varphi_R)}\right) .
    \end{equation}
    
    Based on \eqref{eq:R_t}, and using the same method as in Section \ref{subsection:classic_model} for the monostatic case, it is now possible to derive a model for bistatic returns of a rotor with $N_B$ rotating blades of length $L_B$ each.

\subsection{OFDM Waveform}\label{subsection:OFDM}

   The model from \eqref{eq:base_radar_signal} assumes a simple CW waveform with center frequency $f_0$, where $\omega_0 = 2 \pi f_0$. 
   This formulation can be considered a good approximation for radars with narrow bandwidth. 
   However, for broadband systems this formulation might not properly reproduce the signal features, since the same target scatter point can produce different Doppler contributions due to the different frequency components of the transmitted signal.

    In ISAC, it is common to use OFDM waveforms and, for some applications, these waveforms can occupy a considerably broad bandwidth. This case deserves a special attention and will be analyzed in the next subsection.

\subsection{Bistatic OFDM Mathematical Formulation}\label{subsection:Model}

    The $m^{th}$ transmitted symbol of a system with $N$ OFDM subcarriers can be written as        
    \begin{equation}\label{eq:OFDM_tx_1}
        x(\mu,m) = \sum_{n=1}^{N} D(n,m)\exp\left(j \omega_n (m + \frac{\mu}{N})T\right),
    \end{equation}
where $T$ is the time duration of the OFDM symbol, and $\mu$ and  $m$ are discrete fast-time and slow-time indexes, respectively, so that the total discrete time $t = mT + \frac{\mu}{N}T$. 
    
    Additionally, ${D(n,m)}$ is the complex amplitude for each subcarrier, also called \textit{complex modulation symbol}, given by the modulation technique, e.g. phase-shift keying (PSK). 
    
    Furthermore, in order to have orthogonality, $\omega_n$ is given by
    \begin{equation}
        \omega_n = 2 \pi f_n = 2 \pi (f_0 + \frac{n}{T}),
    \end{equation}
    with $f_n$ denoting the central frequency of each subcarrier.

    So, as $\exp(j2\pi n m) = 1$, and using $t$ in the expression as a matter of simplicity, (\eqref{eq:OFDM_tx_1}) can be rewritten as
    \begin{equation}\label{eq:OFDM_tx_2}
        x(\mu,m) = \exp(j \omega_0 t) \sum_{n=1}^{N} D(n,m)\exp(j 2\pi \frac{n\mu}{N}) .
    \end{equation}        

    Then, using the same logic as in \cite{Qu2019},  the HRR baseband bistatic returns from a scatter point $P$, with respect to the $m^{th}$ symbol, can be written as
    
    \begin{align*}\label{eq:OFDM_ry_P}
        y_P^{(b)}(t, \tau) = &\sum_{n=1}^{N}\Bigg\{ \gamma_n D(n,m)\exp \bigg[j \omega_n \left(t-\frac{R_P^{(b)}(t)}{c}\right) \bigg]\\& \cdot \operatorname{rect} \left(\frac{\tau-\frac{R_P^{(b)}(t)}{c}}{T} - \frac{1}{2} \right)\Bigg\} + z(t)\numberthis . 
    \end{align*}

\begin{figure}[t]
    \centering
    \includegraphics[width=0.45\textwidth, height=2in]{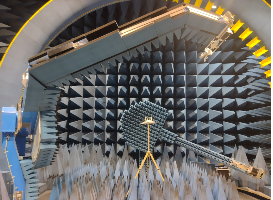}
    \caption{BiRa measurement system: the two gantries can scan the object at the center of a turntable more than the dimension of the hemisphere for both incident and scattering.}
    \label{fig_DJI}
\end{figure}

    Now, if $P$ is a rotating scatter point, the bistatic range $R_P^{(b)}(t)$ is given by (\eqref{eq:R_t}).

    Once again, considering a blade $i$ as a line of infinite point scatters, each of its returns can be calculated by using

    \begin{align*}\label{eq:OFDM_rx_l}
        y_i^{(b)}(t, \tau) =& \int_{0}^{L_B} \sum_{n=1}^{N} \Bigg\{ \gamma_n D(n,m)\exp \bigg[j \omega_n \left(t-\frac{R_P^{(b)}(t)}{c}\right) \bigg]\\& \cdot \operatorname{rect} \left(\frac{\tau-\frac{R_P^{(b)}(t)}{c}}{T} - \frac{1}{2} \right) \Bigg\} \, dl + z(t). \numberthis
    \end{align*}

    If $l_1$ and $l_2$ correspond to the limits of the blade length comprised by the pulse length in the delay $\tau = \frac{\mu}{N}T$, then
    
    \begin{equation}\label{eq:OFDM_rx_l_2}
    \begin{split}
        y_i^{(b)}(t, \tau) =& \sum_{n=1}^{N}\Bigg\{ \gamma_n D(n,m)\exp \left(j \omega_n t \right) \\&\cdot\int_{l_1}^{l_2}{\exp \bigg[-j \frac{\omega_n}{c} (R_P^{(b)}(t)) \bigg]}{dl}\Bigg\} + z(t).
    \end{split}
    \end{equation}

    The returns of all $N_B$ blades, with $\varphi_B$ given, similarly to \eqref{eq:phi_B_i}, by $\varphi_B(i) = \varphi_B + 2\pi\frac{i-1}{N_B}$, can be summed with
    
    \begin{equation}\label{eq:OFDM_rx_l_3}
        y^{(b)}(t, \tau) = \sum_{i=1}^{N_B}\Bigg\{ y_i^{(b)}(t, \tau) \Bigg\}.
    \end{equation}

    Solving the integral, calculating its limits, and eliminating the carrier frequency component, the proposed model is concluded by describing the baseband signal as
    \begin{equation}\label{eq:OFDM_rx_l_5}
    \begin{split}
        &y^{(b)}(\mu, m) = \sum_{n=1}^{N} \Bigg\{ D(n,m)\exp \left(j 2 \pi \frac{n \mu}{N} \right) 
        \\& \cdot\sum_{i=1}^{N_B} \biggl[ \gamma_{ni}(m)e^{j\frac{\omega_n}{c}\left(-R_O(t)+\Delta R_+(\mu, i)\right)}
        \\&\cdot\frac{l_2(\mu)-l_1(\mu)}{2} \operatorname{sinc}\left(\frac{\omega_n}{c} \Delta R_-(\mu, i)\right)\biggr] \Bigg\} + z(t),
    \end{split}
    \end{equation}
    where 
    \begin{equation}
        \Delta R_+(\mu, i) = A_B \frac{l_2(\mu)+l_1(\mu)}{2} \cos(\omega t + \varphi_B(i)),
    \end{equation}
    
    \begin{equation}
        \Delta R_-(\mu, i) = A_B \frac{l_2(\mu)-l_1(\mu)}{2} \cos(\omega t + \varphi_B(i)),
    \end{equation}    
    with $l_1(\mu)$ and $l_2(\mu)$ being the medians, respectively, of the following sets:
        
    \begin{equation}\label{eq:l1_OFDM_2}
        \bigg\{0, \frac{R_O(t) - (\frac{\mu}{N}-1) cT}{A_B \cos(\omega t+\varphi_B(i))}, \frac{R_O(t) - \frac{\mu}{N}cT}{A_B \cos(\omega t+\varphi_B(i))}\bigg\},
    \end{equation}
    
    \begin{equation}\label{eq:l2_OFDM_2}
        \bigg\{L_B, \frac{R_O(t) - (\frac{\mu}{N}-1) c T}{A_B \cos(\omega t+\varphi_B(i))}, \frac{R_O(t) - \frac{\mu}{N} c T}{A_B \cos(\omega t + \varphi_B(i))}\bigg\}.
    \end{equation}

    Micro-Doppler signatures, generated by the proposed model, are presented and compared to measurement in Section \ref{section:Comparison}.

\begin{table}[t]
    \caption{MICRO-DOPPLER MEASUREMENT SETUPS}
    \centering
    \begin{tabular}{|c|l|l|l|}
        \hline
        \textbf{Parameter} & \textbf{Setup 1} & \textbf{Setup 2} \\
        \hline
        Waveform & OFDM-based & OFDM-based \\
        \hline
        Central frequency & 3.7 GHz & 7 GHz \\
        \hline
        Bandwidth & 200 MHz & 2.46 GHz \\
        \hline
        Total number of carriers & 1600 & 2500 \\
        \hline
        Carriers with energy & 1280 & 2048 \\
        \hline
        Symbol duration & $8 \mu s$ & $1.02 \mu s$ \\
        \hline
        Rotating propellers & 1 & 2 \\
        \hline
        Blades per propeller & 2 & 2 \\
        \hline
        Propeller radius & 16.55 cm & 16.55 cm \\
        \hline
        Rotation speed & 1500 rpm & 1500 and 2000 rpm \\
        \hline
        Propeller material & carbon fiber & carbon fiber \\
        \hline
    \end{tabular}
    \label{tab_MicroDoppler_measurement_setup}
\end{table}

\begin{figure}[b]
    \centering
    \includegraphics[width=0.45\textwidth, height=2in]{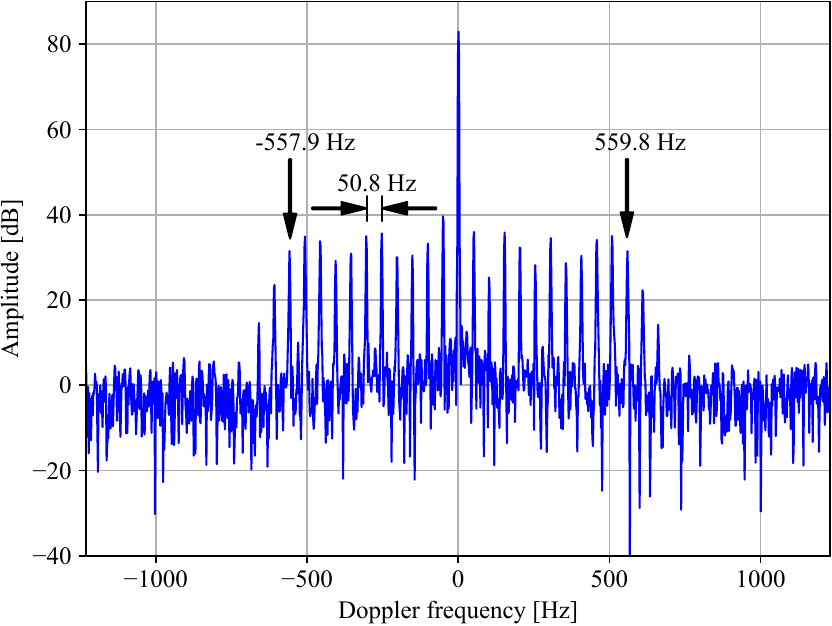}
    \caption{Rotating propeller Micro-Doppler measurement signature in frequency domain (16384 symbols, subsampling = 8).}
    \label{fig_FFT_copter}
\end{figure}


\vspace{5pt}
\section{BiRa Measurements}\label{section:BiRa}
\subsection{BiRa Measurement System and Drones} \label{subsection:BiRa}

In order to validate the model from Section \ref{section:Model}, it is necessary to perform measurements combining both bistatic geometry and broadband OFDM waveform.
A solution for this kind of measurement is the recently inaugurated BiRa measurement system, situated at the Thuringian Innovation Center for Mobility (ThIMo) of the Ilmenau University of Technology. 

This advanced system comprises two pivoting gantries, each equipped with a transmitting (Tx) and receiving (Rx) antenna, respectively \cite{Thoma2}. 
The target object of interest is positioned at the center of the turntable as in Fig. \ref{fig_DJI}.
Therefore, the target can be illuminated and observed from any desired bistatic angle in azimuth, from any elevation bistatic angle in the upper hemisphere, and from 25° elevation in the lower hemisphere.

A particularly noteworthy feature is that its mechanical system is independent and can be integrated with a vast variety of Tx/Rx measurement systems, thus allowing multiple combinations of applications, waveforms, bandwidths, carrier frequencies, etc. 
Furthermore, BiRa has a wideband channel sounder, which can be used associated to the software-defined radio (SDR) system architecture described in \cite{rfsoc_paper}, allowing high range resolution (HRR) micro-Doppler measurements, with instantaneous bandwidth larger than 2 GHz.

Moreover, BiRa allows efficiently setting up sets of Tx/Rx positions, with joint control of gantries positions, turntable azimuth, and device under test (DUT) settings. 
Therefore, it is possible to perform long fully automated measurement runs, including multiple bistatic angles, without requiring any operator interaction.

\begin{figure}[t]
\centering
    \includegraphics[width=3.0in, height=1.7in]{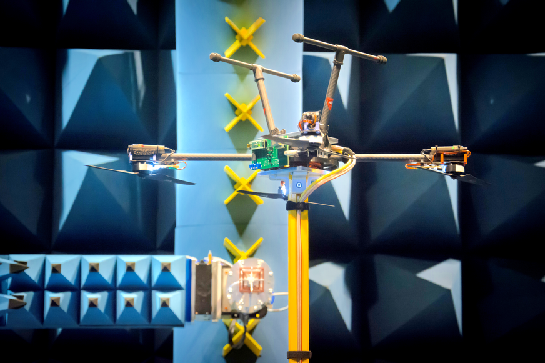}
    \caption{Micro-Doppler measurement in BiRa.}
    \label{fig_mockup}
\end{figure}

Another important aspect for this measurement infrastructure, is that for micro-Doppler analysis purposes, in order to provide reference speed, it is desirable to have full control over the rotors' rotation, which is not feasible with most standard commercial drones.
Therefore, a custom-built system was assembled, allowing us to remotely set the propeller's angular velocities. 
This system, presented on Fig. \ref{fig_mockup}, is composed by a Tarot IRON MAN 650 mechanical structure, with motors, sensors, and an Arduino on a PCB board for setting up motors thrust values and measuring rotation speeds, as well as an Ethernet shield for remote commanding.

\begin{figure}[b]
    \centering
    \includegraphics[width=0.45\textwidth, height=2in]{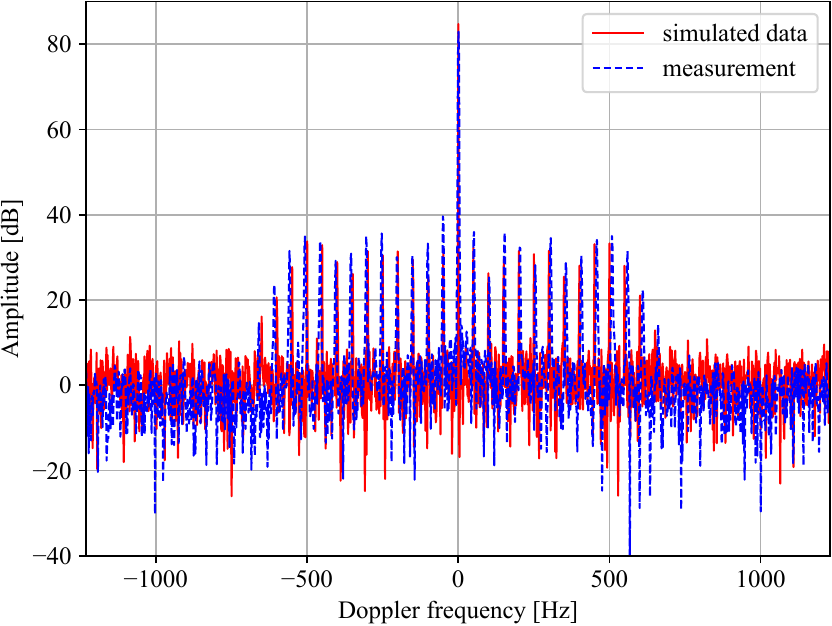}
    \caption{Comparison between simulation and measurement micro-Doppler signatures of a rotating propeller (16384 symbols, subsampling = 8).}
    \label{fig_simul_measurement}
\end{figure}

\begin{figure}[t]
    \centering
    \includegraphics[width=0.45\textwidth, height=2.5in]{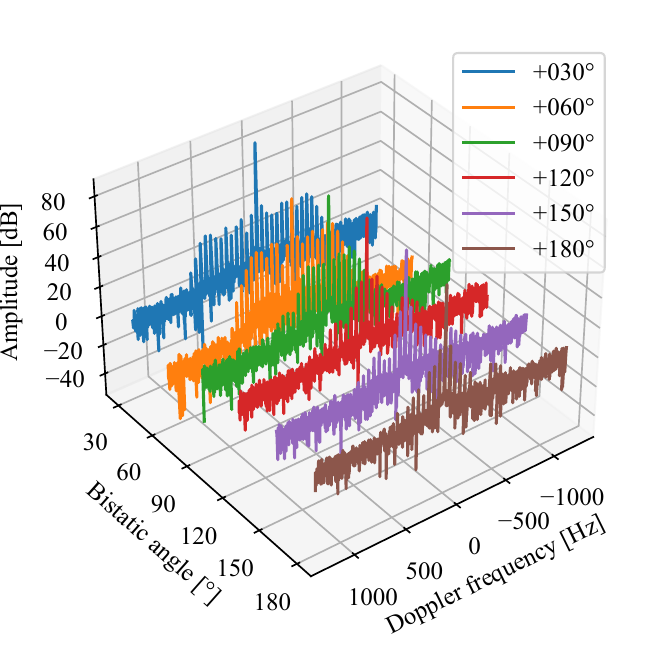}
    \caption{Comparison of frequency-domain micro-Doppler signatures of a rotating propeller from measurement data, with different bistatic angles (16384 symbols, subsampling = 8). The legend indicates the bistatic angle in each measurement.}
    \label{fig_bistatic_angles}
\end{figure}

\subsection{Bistatic Micro-Doppler Measurement Setup} \label{subsection:BiRa_measurement}
    Two setups were implemented, using a wideband OFDM-based transmit signal called Newman sequence with constant spectral magnitude and minimal crest-factor \cite{Boyd1986}. 
    Table \ref{tab_MicroDoppler_measurement_setup} details the measurement system setups.
    
    For micro-Doppler analysis, an OFDM radar processing was performed, as in \cite{braun2014ofdm}, with a back to back measurement as reference for proper measurement calibration.
    The signal was then processed in slow-time, i.e., a vector of the returns of the target in a single range bin, along different symbols. 
    
    Some subsampling can also be applied in order to reduce the amount of data, while focusing the analysis on the bandwidth of interest. 
    In this work, a subsampling factor of 8 means that we use only every eighth symbol available.

\subsection{Measurements Analysis} \label{subsection:measurement_analysis}
    A Doppler frequency signature of the custom drone with only one propeller rotating, obtained from the measurement, is presented in Fig. \ref{fig_FFT_copter}. 
    There, we can see, above the noisy basis, a strong impulse centered in $f = 0$ Hz, and many impulses on both sides of this stronger one. 
    The central impulse occurs due to reflections of the drone's static parts, while the other impulses come from the rotating parts. 
    These impulses are separated by $\Delta f = N f_{rot}$, where $f_{rot}$ is the propeller rotation frequency and $N$ is the number of blades in the propeller. 
    In this case, the rotation frequency was $f_{rot} = 1500 \text{ rpm} = 25$ Hz and $N = 2$, so $\Delta f = 50$ Hz, which is consistent with what can be seen in the image. 
    
    We can also see that this sequence of impulses ends in the Doppler frequencies equivalent to the target's maximum positive and negative relative speeds. 
    In the current measurement, the maximum velocity is given by $v = 2\pi f_{rot} l = 2\pi \times25\times0.1655 = 26$ m/s, where $l = 0.1655$ m is the radius of the propeller. 
    Bistatic Doppler is given by \cite{willis2005bistatic}:
    
    \begin{equation}\label{equ_fd}
        f_D=\frac{2 v \cos(\beta/2) \cos(\delta)}{\lambda}
    \end{equation}
    where $\beta$ is the bistatic angle and $\delta$ the angle between the bistatic bisector, and $v$ the direction of target motion. 
    In this case, we have $\delta$ = 0° and $\beta$ = 60°. So ${f_D}_{max}$= 555.4 Hz.
    
    An important effect that can be inferred from equation \eqref{equ_fd} is the dependency that the Doppler spread has on the bistatic angle $\beta$. Fig. \ref{fig_bistatic_angles} shows how the Doppler spread due to the propeller rotation decreases when the bistatic angle increases. 
    As expected, the maximum Doppler spread is observed in the monostatic case, while in forward configuration almost no Doppler spread is present.
    We can also notice, that distance among impulses does not change with respect to bistatic angles, since they depend only on the rotation frequency and the number of blades.


\vspace{5pt}
\section{Comparison Between Simulation and Measurement}\label{section:Comparison}
\subsection{Low Resolution Case}
    Fig. \ref{fig_simul_measurement} presents a comparison between two micro-Doppler signatures: one simulated using the model developed in Section \ref{section:Model}, and one obtained from the measurement described in Section \ref{section:BiRa}. Both use the parameters described in Table \ref{tab_MicroDoppler_measurement_setup} Setup 1, and geometrical configuration with bistatic angles of 60°. 
    
    Additionally, to generate the simulated data, the OFDM coefficients $D(n,m)$ were obtained from the reference data used for the measurement.
    
    Fig. \ref{fig_simul_measurement} shows that the simulated signature presents the impulses mentioned in Section \ref{subsection:measurement_analysis} in the same positions as in the measurement data target signature.
    
    \begin{figure}[t]
        \centering
        \includegraphics[width=0.45\textwidth, height=2.5in]{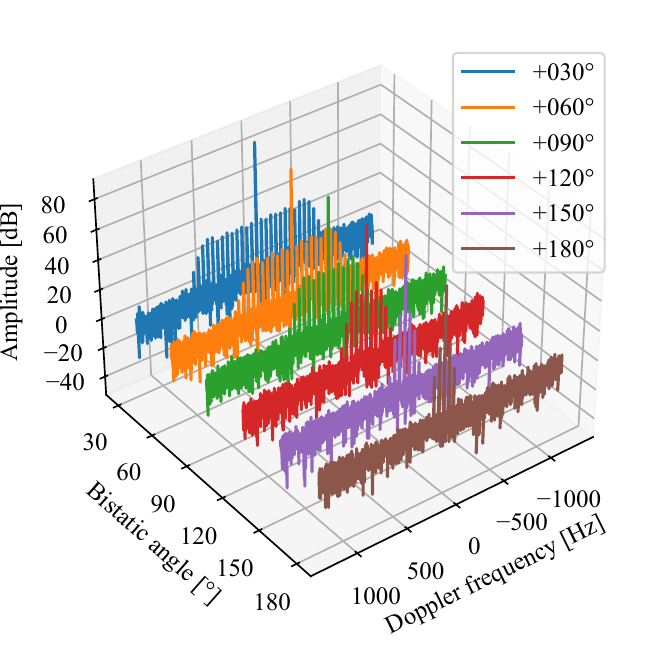}
        \caption{Comparison of frequency-domain micro-Doppler signatures of a rotating propeller from simulated data, with different bistatic angles (16384 symbols, subsampling = 8). The legend indicates the bistatic angle in each simulation.}
        \label{fig_bistatic_angles_simul}
    \end{figure}

    The simulated signature well matches the response obtained from measured data, presenting the same Doppler spread and the same distance between impulses, which are the most important features for classification.

    Comparison between Fig. \ref{fig_bistatic_angles_simul} and Fig. \ref{fig_bistatic_angles} shows that the proposed model is also capable of emulating the dependence on the bistatic angle seen in the measurement data. 
    
    Finally, the Pearson correlation coefficient, i.e. the ratio between the covariance of the simulated and the measured micro-Doppler signatures, and the product of their standard deviations, returns a cross-correlation coefficient of approximately 0.98 across the bistatic angles from 30° to 180°.

\subsection{High Range Resolution Case} 
    The classical model for rotating blades micro-Doppler presented in \cite{Jianjiang2005} assumes that the target response is completely contained in one range resolution cell.
    
    However, when the range resolution is smaller than the target dimensions, the latter has an extended response in fast-range dimension, which stands for the high range resolution (HRR) case.
    This is achieved by using higher signal bandwidth $B$, since the range resolution is given by $\Delta R = \frac{c}{2B}$. \cite{Richards2014}
    
    An additional advantage of the model proposed in the current work is the ability to generate HRR data.

    It is also possible to simulate a drone with multiple propellers by setting the coordinates of the rotation center point \textbf{O} (Fig. \ref{fig:rot_pt1}) of each rotor according to the drone geometry, generating each propeller returns using the proposed model, and then summing the returns from each propeller and from the static parts of the drone.    
    
\begin{figure}[t]
    \centering
    \includegraphics[width=0.5\textwidth, height=1.5in]{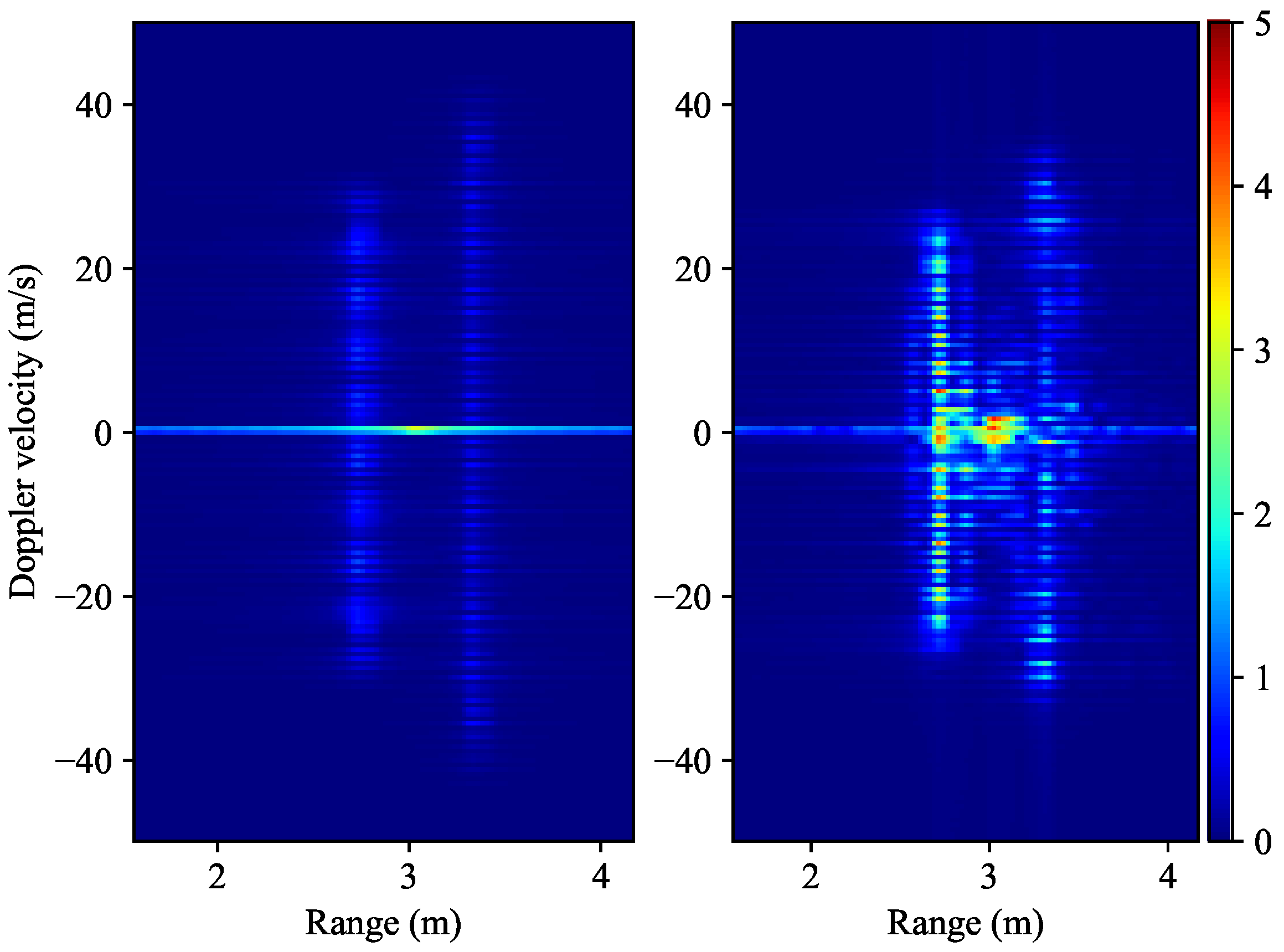}
    \caption{Comparison of simulation (left) and measurement (right) range-Doppler signatures of a drone with two rotating propellers.}
    \label{fig_measurement_HRR}
\end{figure}

    Fig. \ref{fig_measurement_HRR} shows a comparison between the range-Doppler signatures of a drone with two propellers from two sources: simulated from the proposed model and retrieved from BiRa measurements. 
    Both cases use the parameters from Table \ref{tab_MicroDoppler_measurement_setup} Setup 2, thus represent a high range resolution scenario. 
    The main properties of the drone micro-Doppler signatures for the sake of classification, such as number of propellers, position and Doppler spread of each propeller, and distance between impulses in frequency direction are the same on both sides.
    In order to improve intensity match between simulation and measurement, a better RCS model for propellers and static parts returns can be included, by adjusting $\sigma_P$ in \eqref{eq:radar_equation}.

\subsection{Advantages and Disadvantages}

    Many other methods can be used for modeling drone propellers, such as Finite Element Method (FEM), ray tracing, primitive-based modeling from finite discrete scatter points, and classic primitive continuous modeling.
    Compared to other methods, the most important advantages of the proposed method are its simplicity, low computational cost, and ability to represent the HRR case, while taking the waveform into account.
    Among the disadvantages, it does not take into account some electromagnetic phenomena, such as diffraction and multipath, and it also cannot reproduce more complex phenomena such as frequency components due to vibration.
    However, its simplicity makes it suitable for applications where a large amount of simulated data with various parameters is required.


\vspace{5pt}
\section{Conclusions and Outlook}\label{section:Conclusion}
This paper presents a model for simulating the bistatic micro-Doppler signature of drone propellers in distributed ISAC. 
The simulated target signatures are compared to signatures obtained from real BiRa measurement. 

As a conclusion, it is shown that the proposed model produces the same basic features of the bistatic micro-Doppler signature achieved in the signatures obtained from measurement data. In addition, the model reproduces the relation between the bistatic angle and the micro-Doppler spread in the same way it appears in the measurement data signature.

Future work can comprise the addition of a more realistic reflectivity model for the main body of the target in the HRR case, the use of simulated data using this model to train target classification algorithms, and the validation of this training method using measurement data.

\section*{Acknowledgment}
The authors would like to thank Dr.Ing. Tobias Nowack and M.Sc. Masoumeh Pourjafarian for their support throughout the BiRa measurement in ThiMo.
This research is funded by the BMBF project 6G-ICAS4Mobility, Project No.16KISK241, by the Federal State of Thuringia, Germany, and by the European Social Fund (ESF) under grants 2017 FGI 0007 (project "BiRa") and 2021 FGI 0007 (project "Kreatör").

\bibliographystyle{IEEEtran}
\bibliography{references}
\vspace{12pt}

\end{document}